\pdfoutput=1

\documentclass[11pt,twoside,a4paper,cmspaper,final,collab]{cms-tdr}
\def\svnVersion{a6927b3-D}\def\svnDate{2020/04/25}\def\cmsCernNoTag{CERN-EP-2019-229}\def\cmsCernDate{\today}\def\cmsMessage{"Published in Physics Letters B as \href{http://dx.doi.org/10.1016/j.physletb.2020.135448}{\doi{10.1016/j.physletb.2020.135448}.}"}

\begin{document}\cmsNoteHeader{EXO-19-004}

\hyphenation{had-ron-i-za-tion}
\hyphenation{cal-or-i-me-ter}
\hyphenation{de-vices}
\newcommand{\CaloLumi}{{18.3\fbinv}}
\newcommand{\minFitMass}{290\xspace}
\newcommand{\maxFitMass}{1000\xspace}
\newcommand{\resMassLow}{350\xspace}
\newcommand{\resMassHigh}{700\xspace}
\newcommand{\DeltaEtaCut}{1.1\xspace}
\newcommand{\WideJetCone}{1.1\xspace}
\newcommand{\ptThreshold}{30\xspace}

\newcommand{\couplingLimitLow}{0.10\xspace}
\newcommand{\couplingLimitHigh}{0.15\xspace}

\newcommand{\SM}{\PQq}
\newcommand{\DM}{\PGc}
\newcommand{\gDM}{\ensuremath{g_{\text{DM}}}\xspace}
\newcommand{\gq}{\ensuremath{g_{\Pq}}\xspace}
\newcommand{\gqp}{\ensuremath{g'_{\Pq}}\xspace}
\newcommand{\mjj}{\ensuremath{m_{\mathrm{jj}}}\xspace}

\cmsNoteHeader{EXO-19-004}
\title{Search for dijet resonances using events with three jets in proton-proton collisions at $\sqrt{s}=13\TeV$}

\date{\today}

\abstract{
A search for a narrow resonance with a mass between 350 and 700\GeV, and decaying into a pair of jets, is performed using proton-proton collision events containing at least three jets. The data sample corresponds to an integrated luminosity of 18.3\fbinv recorded  at $\sqrt{s}=13\TeV$ with the CMS detector. Data are collected with a technique known as ``data scouting", in which the events are reconstructed, selected, and recorded at a high rate in a compact form by the high-level trigger. The three-jet final state provides sensitivity to lower resonance masses than in previous searches using the data scouting technique. The spectrum of the dijet invariant mass, calculated from the two jets with the largest transverse momenta in the event, is used to search for a resonance. No significant excess over a smoothly falling background is found. Limits at 95\% confidence level are set on the production cross section of a narrow dijet resonance and compared with the cross section of a vector dark matter mediator coupling to dark matter particles and quarks. Translating to a model where the narrow resonance interacts only with quarks, upper limits on this coupling range between 0.10 and 0.15, depending on the resonance mass. These results represent the most stringent upper limits in the mass range between 350 and 450\GeV obtained with a flavor-inclusive dijet resonance search.
}

\hypersetup{
  pdfauthor={CMS Collaboration},
  pdftitle={Search for dijet resonances using events with three jets in proton-proton collisions at sqrt(s) = 13 TeV},
  pdfsubject={CMS},
  pdfkeywords={CMS, physics, dijets, resonances},
}

\maketitle

\section{Introduction}

Many models of new physics predict the existence of new massive particles coupled to quarks.
The production and decay of these particles into two jets, known as dijets, have been searched for since the first high-energy hadron colliders came into operation \cite{Albajar:1988rs, Alitti:1990kw, Alitti:1993pn, Abe:1989gz,Abe:1993it,Abe:1995jz,Abe:1997hm,Aaltonen:2008dn, Abazov:2003tj}.
In some models, these particles act as mediators linking the standard model (SM) to new physics sectors containing dark matter (DM) particle  candidates \cite{Abercrombie:2015wmb, Boveia:2016mrp,Chala:2015ama,Abdallah:2015ter}.
New mediators interacting with both quarks and DM particles (\DM) have been searched for using different methods:
in the direct searches for DM, by measuring the recoil of an SM particle caused by the scattering with a DM particle ($\DM\SM \to \DM\SM$, $t$-channel) \cite{DM-XENON100-1,DM-XENON100-2,DM-XENON1T,DM-PandaXII-1,DM-PandaXII-2,DM-LUX-1,DM-LUX-2,DM-DarkSide,DM-SuperCDMS1,DM-SuperCDMS2,DM-SuperCDMS3};
with astrophysics detectors, by looking for SM particles produced through the annihilation of DM particles ($\DM\DM \to \SM\SM$, $s$-channel) \cite{DM-AMS1,DM-AMS2,DM-AMS3,DM-IceCube,DM-PAMELA1,DM-PAMELA2,DM-DAMA,DM-ICE,DM-FermiLAT1,DM-FermiLAT2,DM-FermiLAT3,DM-SuperKamiokande,DM-HESSE1,DM-HESSE2,DM-HESSE3};
and at hadron colliders, by detecting the momentum imbalance due to the production of DM particles ($\SM\SM \to \DM\DM$, $s$-channel) \cite{DM-ATLAS-1, DM-ATLAS-2, DM-ATLAS-3, DM-ATLAS-4, DM-CMS-1, DM-CMS-2, DM-CMS-3}.
The search for dijet resonances at hadron colliders ($\SM\SM \to \SM\SM$, $s$-channel) can be compared with such DM searches and the results are particularly sensitive for models where the decay of the mediator into DM particles is forbidden for kinematic reasons.
The search for dijet resonances is also sensitive to the signals predicted by other models \cite{Anchordoqui:2008di,Cullen:2000ef,ref_diquark,ref_qstar,Baur:1989kv,ref_axi,Chivukula:2013xla,ref_coloron,Han:2010rf,ref_gauge,ref_rsg}.

Experiments at the CERN LHC have used various techniques to search for resonances in the dijet invariant mass spectrum.
From searches where both jets are individually resolved, the ATLAS and CMS Collaborations have set limits for resonances with masses above 450 and 600\GeV, respectively,  in $\sqrt{s}=13\TeV$ proton-proton collisions \cite{Aaboud:2017yvp,Aaboud:2018fzt, Sirunyan:2018xlo}, and above 250 and 500\GeV, respectively, in 8\TeV collisions \cite{Aad:2014aqa,Khachatryan:2016ecr}.
In the sub-\TeVns mass range, another search by the ATLAS Collaboration at 13\TeV for dijet resonances, produced in association with a photon from initial-state radiation, has set limits in the mass region between 225 and 1100\GeV \cite{Aaboud:2019zxd}.
A search by the CMS Collaboration at 8\TeV for resonances decaying into two bottom quarks, experimentally identified as \PQb-tagged jets, has set limits in the mass range of 325--1200\GeV \cite{Sirunyan:2018pas}.
Finally, the ATLAS and CMS Collaborations have set limits in the mass range below 220 and 450\GeV, respectively, from searches for Lorentz-boosted resonances decaying into a quark-antiquark pair reconstructed as a single jet \cite{ISRATLAS, Sirunyan:CMS-boosted-paper,Sirunyan:2019sgo}.

This paper presents a search for a dijet resonance in three-jet events that is sensitive to narrow resonances with mass between \resMassLow and \resMassHigh\GeV.
The search is based on data from $\Pp\Pp$ collisions at $\sqrt{s} = 13\TeV$ collected in 2016, corresponding to an integrated luminosity of \CaloLumi.
To obtain a large trigger efficiency in the  mass range of $\resMassLow$--\resMassHigh\GeV, we select a three-jet final state and utilize a special high-rate trigger with low jet \pt thresholds.
This trigger uses a technique known as ``data scouting'' described in Section  \ref{sec:trigger-simulation}.
This search is limited to data collected in the year 2016 in order to take advantage of the low trigger thresholds used in that data period.
After 2016, these thresholds were raised in order to limit the trigger rate increase due to the larger instantaneous luminosity and pileup.

\section{The CMS detector}

A detailed description of the CMS detector, together with a definition of the coordinate system used and the relevant kinematic variables, can be found in Ref.~\cite{Chatrchyan:2008zzk}.
The central feature of the CMS apparatus is a superconducting solenoid of 6\unit{m} internal diameter, providing a magnetic field of 3.8\unit{T}. Within the solenoid volume are a silicon pixel and strip tracker, a lead tungstate crystal electromagnetic calorimeter (ECAL), and a brass and scintillator hadron calorimeter (HCAL), each composed of a barrel and two endcap sections. Muons are detected in gas-ionization chambers embedded in the steel flux-return yoke outside the solenoid.

The jets used by this analysis are calorimeter-based jets that are reconstructed from the energy deposits in the calorimeter towers, clustered using the anti-\kt algorithm~\cite{Cacciari:2008gp, Cacciari:2011ma} with a distance parameter of 0.4. In this process, the contribution from each calorimeter tower is assigned a momentum, the absolute value and the direction of which are found from the energy measured in the tower, and the coordinates of the geometrical center of the tower. The raw jet energy is obtained from the sum of the tower energies, and the raw jet momentum from the vectorial sum of the tower momenta. The raw jet energies are then corrected to establish a uniform relative response of the calorimeter in pseudorapidity $\eta$ and a calibrated absolute response in transverse momentum \pt.
The calorimetric jet energy resolution is typically 40\% at a \pt of 10 \GeV, 12\% at 100 \GeV, and 5\% at 1 \TeV, resulting in a calorimetric dijet mass resolution of about 10\% for resonance masses between 350 and 700 \GeV.
Events of interest are selected using a two-tiered trigger system~\cite{Khachatryan:2016bia}. The first level (L1), composed of custom hardware processors, uses information from the calorimeters and muon detectors to select events at a rate of around 100\unit{kHz} within a time interval of less than 4\mus. The second level, known as the high-level trigger (HLT), consists of a farm of processors running a version of the full event reconstruction software optimized for fast processing, and reduces the event rate to around 1\unit{kHz} before data storage.

\section{Data and simulated event samples}
\label{sec:trigger-simulation}

We selected events requiring ${\HT>240\GeV}$ at the L1 trigger and $\HT>250\GeV$ at the HLT, where \HT is the scalar \pt sum of jets with $\pt > 40\GeV$ and $\abs{\eta}<2.5$.
The rate of this trigger was about 4\unit{kHz} at an instantaneous luminosity of $1 \times 10^{34} \unit{cm$^{-2}$s$^{-1}$}$.
The amount of data generated by such a high-rate trigger alone using the standard data-taking format would have saturated the computing and storage systems of the CMS experiment.
For this reason, we used a special data-taking technique, which consisted of saving only the calorimeter-based jets reconstructed by the HLT, instead of the full detector readout.
The size of this reduced data format is about 0.5\% of the full event size.
This technique is known as ``data scouting'' and was used in previous CMS dijet resonance searches \cite{Khachatryan:2016ecr,Sirunyan:2018xlo}.

Data scouting allows the analysis of a very large rate of data passing the HLT trigger, only limited by the overall rate of the L1 trigger.
To keep a constant rate, the L1 trigger \HT threshold was raised from 240 to 360\GeV as the instantaneous luminosity increased.
This search is limited to data collected in 2016 with the lower L1 trigger threshold $\HT > 240\GeV$, in order to obtain the maximum sensitivity for low mass resonances.
From a sample of events collected with a minimum-bias trigger and passing the selection discussed
below, we measured a trigger efficiency larger than 99\% for a dijet invariant mass greater than 290\GeV.

Signal events corresponding to a narrow vector resonance decaying into quark-antiquark pairs  were generated using  the \MGvATNLO version 2.2.2 generator  at leading order \cite{Alwall:2014hca,MadDM},
with the \PYTHIA 8.205 generator \cite{Sjostrand:2014zea} incorporating the CUETP8M1 underlying event tune \cite{Khachatryan:2015pea} providing the description of fragmentation and hadronization.
The generated resonance width is negligible compared to the experimental dijet mass resolution, which is about 10\%.
The detailed simulation of the CMS detector response is performed using the \GEANTfour package \cite{GEANT4}.
The simulated signal events include multiple overlapping $\Pp\Pp$ interactions per bunch crossing (pileup) as observed in the data.
Additionally, to provide a framework for interpreting the results in terms of a DM mediator, signal cross sections were computed at leading order with \MADGRAPH for a vector boson decaying into a quark-antiquark pair, with coupling to quarks $\gq=0.25$, coupling to DM particles $\gDM = 1.0 $, and the mass of DM particles 1\GeV. The NNPDF2.3LO~\cite{Ball:2012cx} parton distribution functions were used.

\section{Event reconstruction and selection}

The discriminating variable in this analysis is the invariant mass of the two jets originating from the resonance decay.
This variable is calculated using jets, reconstructed at the HLT from energy deposits in the calorimeter, and passing the selection $\pt> \ptThreshold\GeV$ and $\abs{\eta} <2.5$.
Spurious jets originating from instrumental noise are rejected by requiring each jet to be detected by both ECAL and HCAL, with at least 5\% of the jet energy in each of the two types of calorimeter.
We form ``wide jets", by clustering the jets already reconstructed by the HLT, using the anti-\kt algorithm with a distance parameter of 1.1.
This algorithm improves the dijet mass resolution and the resonance search sensitivity, by recombining jets from hard final-state radiation to obtain a reduced number of wide jets.
A similar algorithm using a merging distance  of ${\DR = \sqrt{\smash[b]{ (\Delta \eta)^2 + (\Delta \varphi)^2}}< \WideJetCone}$ was employed in previous CMS searches \cite{Sirunyan:2018xlo,Khachatryan:2016ecr}, but it only reconstructed two wide jets per event.
The wide-jet calibrations for the 2016 data scouting sample were already obtained in the low-mass dijet search of Ref. \cite{Sirunyan:2018xlo}, and therefore we apply the same calibrations.
We require at least three wide jets, each with $\pt>72\GeV$, in order to select events that have large \HT and pass the trigger selections.
This requirement is particularly effective in selecting events with low dijet invariant mass, which would be rejected if only two jets were required.
Applying a common threshold to the \pt of the three jets enabled us to minimize the value of the lowest resonance mass to which we are sensitive.
The \pt threshold of the three-jet selection has been chosen with a method that is explained in the next section.
Finally, the two leading wide jets are required to have  $\abs{\eta_1 - \eta_2} < \DeltaEtaCut$ to reduce the quantum chromodynamics multijet background, which is dominated by $t$-channel production of jets.

Since we require at least three wide jets in the event, there are multiple ways to select the dijet system, \ie, the pair of wide jets originating from the resonance decay.
We select as the dijet the two wide jets with the largest and the next-to-largest \pt in the event.
This selection is correct in 70\,(50)\% of simulated signal events with a resonance mass of 700 (350)\GeV.
Wrong combinations arise because either an energetic initial-state radiation jet is included in the dijet selection, or an energetic jet from final-state radiation is emitted with a distance $\DR > 1.1$ from the leading jets and therefore excluded from the reconstruction of the two leading wide jets.
We investigated alternative criteria to select the dijet, such as choosing the jet pair with the largest norm of the vectorial sum \pt.
We found that such alternative criteria do have better performance if the resonance \pt is greater than half the mass, but worse performance for this search.
This is because, for accepted events, the \pt of the resonance is about 150\GeV, which is less than half the resonance mass considered in this search.

\section{Dijet mass spectrum fit}
\label{sec:fit}

\begin{figure}[tbh]\centering
\includegraphics[width=0.45\textwidth]{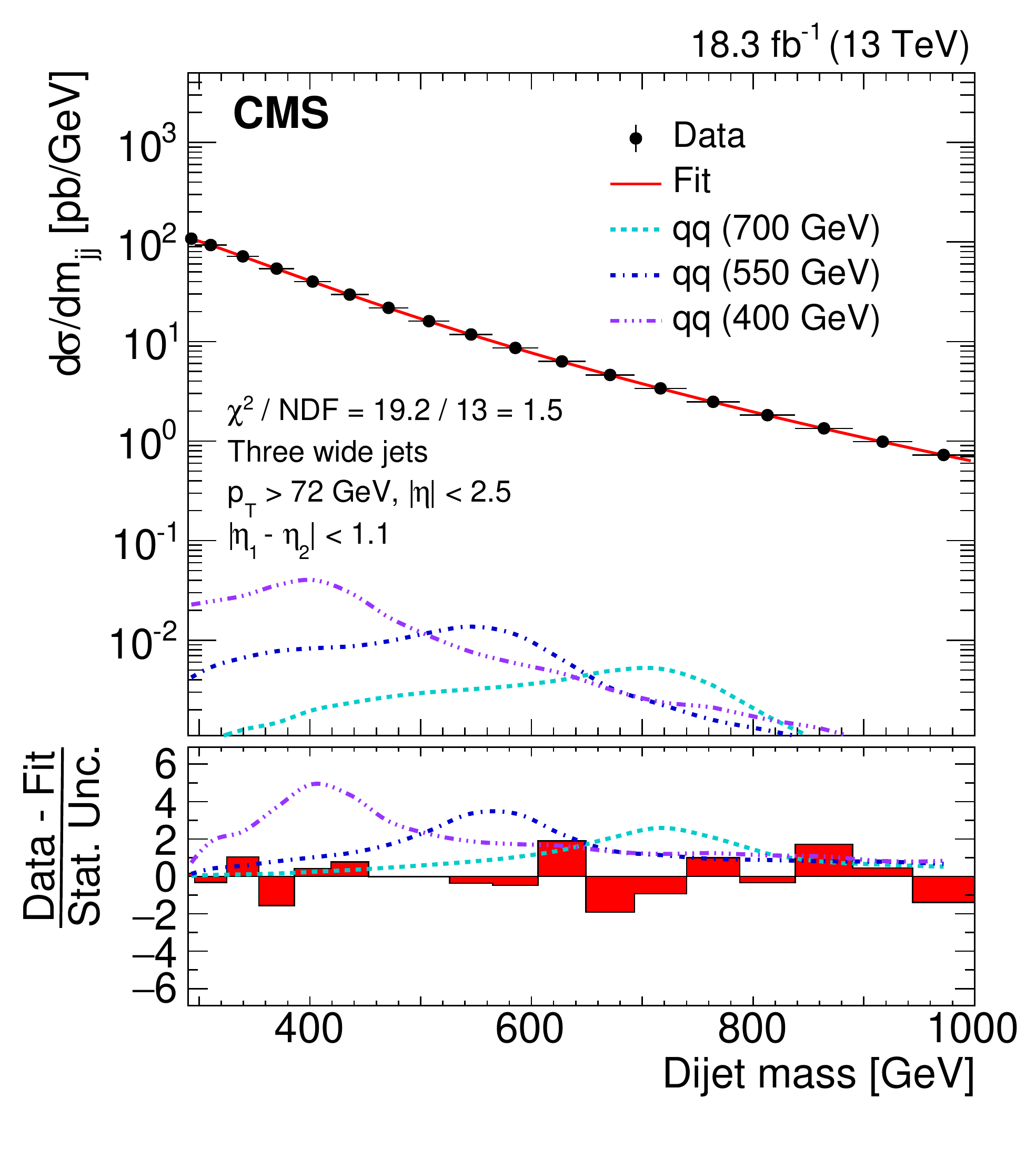}
\caption{\label{fig:djmassHLT} Dijet mass spectrum (points) compared to a fitted parameterization of the background (solid curve).  
The background fit is performed in the range $\minFitMass <\mjj < \maxFitMass\GeV$.
The horizontal bars show the widths of each bin in dijet mass.
The dashed lines represent the dijet mass distribution from 400, 550, and 700\GeV  resonance signals expected to be excluded at 95\% \CL by this analysis.
The lower panel shows the difference between the data and the fitted parametrization, divided by the statistical uncertainty of the data.
}
\label{fig:dijet}
\end{figure}

Figure \ref{fig:dijet} shows the dijet mass ($\mjj$) spectrum.
The background is modeled with the following analytic function,
\begin{equation}
 \frac{\rd \sigma}{\rd \mjj} =   \frac{ p_0 (p_2 x - 1)}{x^{p_1 + p_3 \log{x} + p_4 \log^2{x}}},
 \label{eq:fit}
\end{equation}
where $x$ is defined as $\mjj/\sqrt{s}$, and $p_0$, $p_1$, $p_2$, $p_3$, and $p_4$ are free parameters of the fit.
This function is similar to that used by previous dijet searches  \cite{Sirunyan:2018xlo,Aaboud:2017yvp,Aaboud:2018fzt,Khachatryan:2016ecr,Aad:2014aqa}, with a modification to the numerator.
The new parameterization better fits  the shape of the dijet mass spectrum for three-jet events, which includes the effect of a small inefficiency to pass the trigger for events at the lowest values of dijet mass.
The function has been chosen from a pool of functions using a Fisher test \cite{ref:ftest} with a $95\%$ confidence level (\CL).
The pool of functions is obtained by changing the number of degrees of freedom of the polylogarithmic function in the exponent of the denominator of Eq. (\ref{eq:fit}).
We perform a maximum likelihood fit of the function in Eq. (\ref{eq:fit}) to our data in the mass range ${\minFitMass <\mjj < \maxFitMass\GeV}$.
The chi-square per number of degrees of freedom of the fit is $\chi^2 / \mathrm{NDF} = 19.3/13$, corresponding to a p-value of $0.11$.
Figure \ref{fig:dijet} also shows the expected dijet mass distributions of a resonance signal for three different values of resonance mass.
The data distribution is well modeled by the background parameterization and there is no evidence for a dijet resonance.

The dijet mass bin widths in Fig. \ref{fig:dijet} are the same as in the previous dijet searches, except for the first bin which is more narrow, starting at a dijet mass value of 290\GeV.
This lower bound of the fit range and the jet \pt threshold for the three-jet selection are determined in the following way.
We measure the distribution of the dijet mass in a signal-depleted region defined by replacing the requirement ${\abs{\eta_1 - \eta_2}<1.1}$ with the requirement  ${\abs{\eta_1 + \eta_2}<1.1}$.
The dijet mass in the signal-depleted region is calculated after flipping the sign of $\eta$ of the second jet---the sign of the $z$ component of the momentum of the subleading jet is reversed and then the dijet mass is calculated.
For background events, the dijet mass distribution in the signal-depleted region, so calculated, is closely similar to the dijet mass distribution in the signal region because the variables $\eta_1 - \eta_2$ in the signal region and $\eta_1 + \eta_2$ in the signal-depleted region have approximately the same uniform distribution between $-1.1$ and 1.1.
The signal-depleted region contains about the same number of background events and 50\% fewer signal events, and 35\% of the observed events in the signal-depleted region are also in the signal region.
Small data-driven corrections, which change the observed number of events by less than 5\%, are applied to the dijet mass distribution in the signal-depleted region to make it the same as the background distribution in the signal region.
These corrections, which are applied as a function of the product of the two largest values of jet $\pt$ in the event, are obtained by fitting an analytic function describing this product to the ratio of the numbers of events passing the signal selection to the number of events passing the signal-depleted selection.
The lower edge of dijet mass included in the search, 290\GeV, has been chosen to be the lowest value of the corrected dijet mass in the signal-depleted region  for which the fit of the background parameterization has a Kolmogorov--Smirnov (KS) probability \cite{Kolmogorov,Smirnov1,Smirnov2} larger than 33\%.
The \pt threshold of the three-jet selection, 72\GeV, has been chosen to obtain the lowest possible value for the corrected  dijet mass in the signal-depleted region that could be included in the fit and satisfy the same KS test.
We verified that an injected signal with a strength corresponding to the 95\% \CL expected upper limit does not change the choice of the fit range and the three-jet selection.

\section{Systematic uncertainties}

The asymptotic approximation~\cite{Cowan:2010js} of the modified frequentist \CLs method~\cite{Junk1999,bib-cls} is utilized to set upper limits on signal cross sections, following the prescription described in Ref.~\cite{ATLAS:1379837}.
We use the profiled likelihood ratio as test statistic.
The likelihood is the product of the Poisson probabilities for each of the bins in Fig. \ref{fig:dijet}.
The expected background yield of each bin is determined from the analytic  function described in Eq. (\ref{eq:fit}).
The five parameters of the analytic function are profiled and their uncertainties from the fit to data are the dominant uncertainties.
The shapes of the dijet mass distributions for signals are obtained from simulations.
The systematic uncertainties affecting the signal shape and normalization have a minor impact and are incorporated into the likelihood function via nuisance parameters with log-normal probability distributions.
We account for the uncertainty of 2\% in the jet energy scale \cite{jec} by shifting the dijet mass of the signal distribution by $\pm 2\%$.
The effect of the jet energy resolution uncertainty is included by varying the width of the signal distribution by $\pm 10\%$ \cite{jec}.
The signal acceptance depends significantly on the presence of a jet from initial-state or final-state radiation. 
We estimated the uncertainty of the simulation related to this dependence by modifying by a factor of two both the renormalization ($\mu_R$) and the factorization scales ($\mu_F$) of the initial-state and final-state radiation using the method described in Ref. \cite{Mrenna:2016sih}.
This uncertainty has a negligible effect on the shape of the dijet mass distribution of the signal, and changes the normalization by 10\%.
The uncertainty in the integrated luminosity is 2.5\% \cite{lumi} and affects directly the signal normalization.
The systematic uncertainty due to the choice of the background function has been estimated by measuring the signal yield in pseudo-data spectra generated using alternative background functions.
The measured cross section in each case is the same as that of the injected signal, and this systematic uncertainty is found to be negligible.
We tested the capability of the alternative functions to fit the multijet background by fitting the signal-depleted region described in Section \ref{sec:fit}.
The systematic uncertainties related to pileup, parton distribution functions, underlying events, and parton shower models are also found to be negligible.

\section{Results}

\begin{figure}[tbh] \centering
\includegraphics[width=0.45\textwidth]{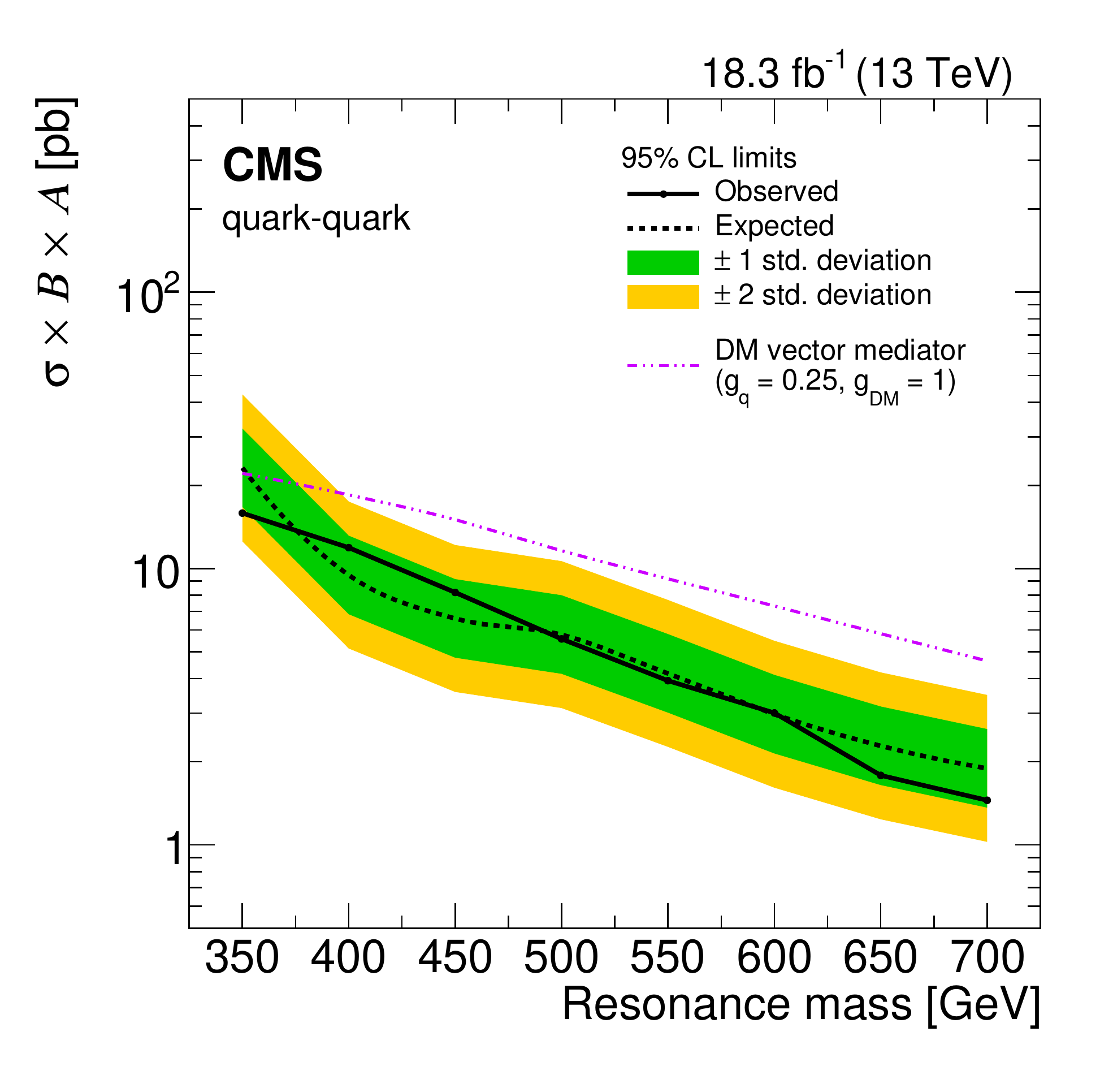}
\caption{Upper limits at 95\% \CL on the product of the cross section, branching fraction, and acceptance as a function of resonance mass for a narrow vector resonance decaying into a pair of quark jets.
The acceptance is calculated for the analysis selection, namely three wide jets with $\pt>72\GeV$ and $\abs{\eta}<2.5$, and ${\abs{\eta_1 - \eta_2} < \DeltaEtaCut} $.
The observed limits (solid curve), expected limits (dashed curve) and their variation at the 1 and 2 standard deviation levels (shaded bands) are shown.
The dashed-dotted curve shows the expected cross section times acceptance for a DM mediator (see text).
}
\label{fig:limits_cross_section}
\end{figure}

\begin{figure}[tbh] \centering
\includegraphics[width=0.45\textwidth]{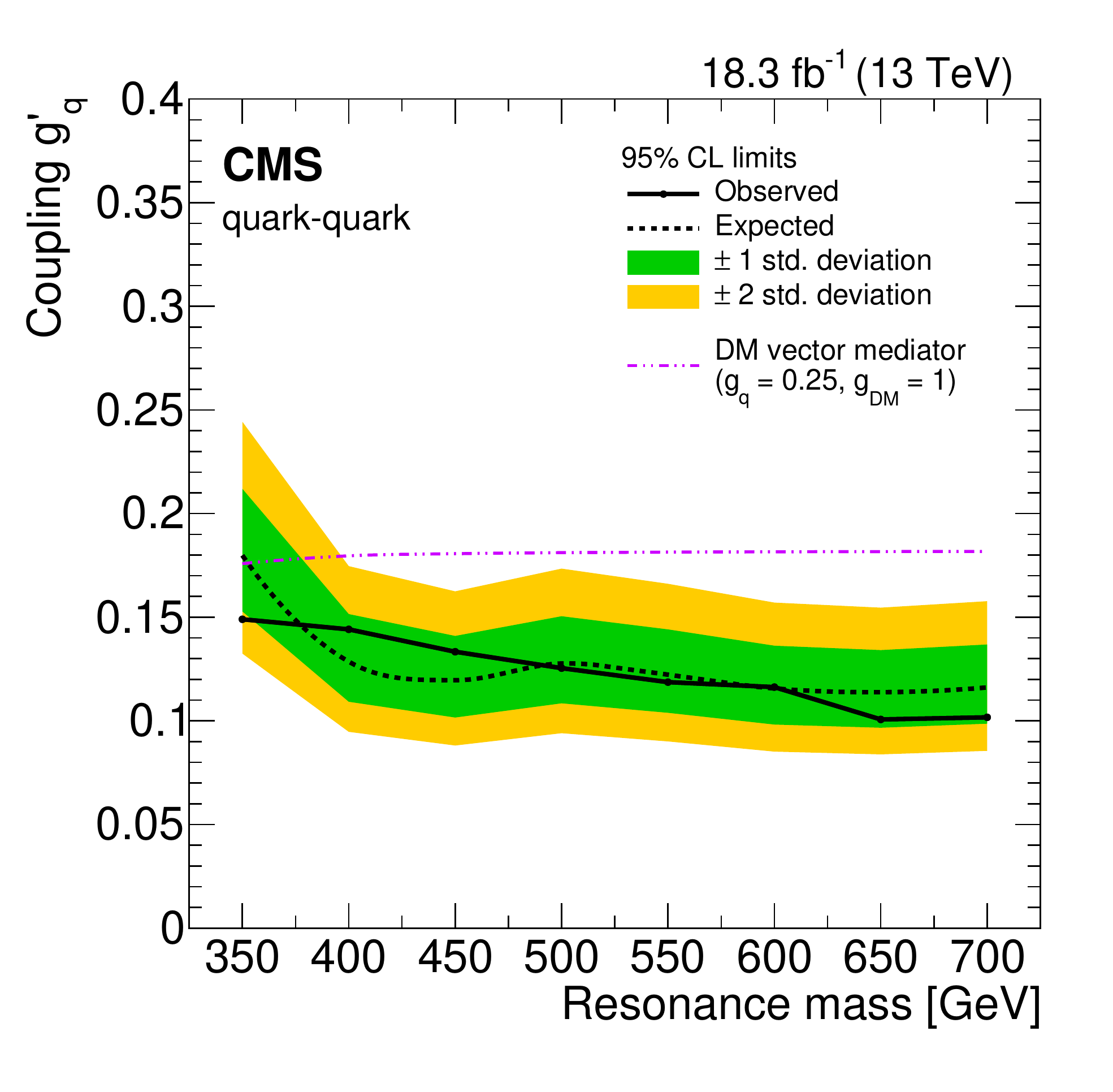}
\caption{Upper limits at 95\% \CL  on the universal quark coupling \gqp,  as a function of resonance mass, for a narrow vector resonance that only couples to quarks.
The observed limits (solid curve), expected limits (dashed curve) and their variation at the 1 and 2 standard deviation levels (shaded bands) are shown.
The dashed-dotted curve shows the coupling strength for which the cross section for dijet production in this model is the same as for a DM mediator (see text).}
\label{fig:limits_couplings}
\end{figure}

Figure \ref{fig:limits_cross_section} shows, as a function of resonance mass, observed and expected upper limits at 95\% \CL on the  product of the cross section, branching fraction, and acceptance  of a narrow vector resonance decaying to jets.
Table \ref{tab:acceptance} shows the acceptance calculated using signal simulations.
\begin{table*}[tbhp]
\topcaption{
Acceptance for a vector resonance  decaying into a dijet as a function of the resonance mass.
The acceptance is calculated using signal simulations for the analysis selection, namely three wide jets with ${\pt>72\GeV}$ and ${\abs{\eta}<2.5}$, and ${\abs{\eta_1 - \eta_2} < \DeltaEtaCut} $.
The errors are dominated by the uncertainty related to the modeling of the jet radiation used in signal simulations.
We estimated this uncertainty by modifying by a factor of two both the renormalization ($\mu_R$) and the factorization scales ($\mu_F$) of the initial-state and final-state radiation \cite{Mrenna:2016sih}.
}
\centering
\begin{tabular}{cccccc}
Resonance mass			& 300\GeV    		& 400\GeV 		& 500\GeV		& 600\GeV		& 800\GeV \\
\hline
Acceptance	& $(4.0 \pm  0.4)\%$	& $(6.7 \pm  0.7)\%$	& $(9.2 \pm  0.9)\%$	& $(10.9 \pm  1.1)\%$	& $(13.6 \pm  1.4)\%$
\label{tab:acceptance}
\end{tabular}
\end{table*}
Limits are presented for resonance masses between 350 and 700\GeV, for which the acceptance of the dijet mass requirement ${\minFitMass <\mjj < \maxFitMass\GeV}$ is large enough to conduct the search.
Figure \ref{fig:limits_couplings} shows that the 95\% \CL upper limits on the coupling \gqp  of a vector resonance that decays only to quarks, defined according to the convention of Ref. \cite{Dobrescu:2013coa}, are between \couplingLimitLow\ and \couplingLimitHigh.
Figures \ref{fig:limits_cross_section} and \ref{fig:limits_couplings} compare the upper limits on the cross section and the coupling \gqp, respectively, with the predictions of a model with a DM mediator that decays to DM particles with masses of 1\GeV, and also decays to quarks.
This analysis excludes a benchmark model of such a DM mediator with coupling to quarks $\gq=0.25$ and coupling to DM particles $\gDM=1$, over the complete mass range 350 to 700\GeV.
In our notation, \gqp is the coupling for a model in which the resonance couples to quarks only, and \gq is the coupling to quarks for a model in which the resonance also couples to DM particles.
We convert \gq into \gqp using the following relationship
\begin{equation} 
\gqp =  \frac{\gq}{ \sqrt{ 1 + 1/ \left( 3 N_{\PQq}(M_\text{med}) \gq^2 \right)}}
\label{g_prime}
\end{equation}
where $N_{\PQq}(M_\text{med})$ is the effective number of quarks
\begin{equation} 
N_{\PQq}(M_\text{med}) = \sum_{q} \left( 1-4\frac{m^2_{q}}{M_\text{med}^2} \right)^{1/2}\left(1+2\frac{m^2_{q}}{M_\text{med}^2}\right)
\label{N_q}
\end{equation}
and the index $q$ runs over the quark flavors (\PQu, \PQd, \PQs, \PQc, \PQb, \PQt) having $m_{q} < M_\text{med}/2$ \cite{Sirunyan:2018xlo, Boveia:2016mrp}.

\section{Summary}

A search for a narrow vector resonance of mass between \resMassLow and \resMassHigh\GeV decaying into two jets has been performed in events containing at least three jets using proton-proton collision data at  $\sqrt{s} = 13\TeV$ at the LHC corresponding to an integrated luminosity of \CaloLumi.
The dijet mass distribution of the two leading jets is smooth, and there is no evidence for a resonance.
Upper limits at 95\% confidence level are set on the product of the cross section, branching fraction, and acceptance as a function of resonance mass.
This search excludes a simplified model of interactions between quarks and dark matter particles of mass 1\GeV, where the interactions are mediated by a vector particle with mass between 350 and 700\GeV, for coupling strengths of $\gq=0.25$ and $\gDM=1$.
Upper limits between \couplingLimitLow\ and \couplingLimitHigh\ are also set on the coupling to quarks \gqp for a vector particle interacting only with quarks.
These results represent the most stringent upper limits in the mass range between 350 and 450\GeV obtained with a flavor-inclusive dijet resonance search.

\begin{acknowledgments}
We congratulate our colleagues in the CERN accelerator departments for the excellent performance of the LHC and thank the technical and administrative staffs at CERN and at other CMS institutes for their contributions to the success of the CMS effort. In addition, we gratefully acknowledge the computing centers and personnel of the Worldwide LHC Computing Grid for delivering so effectively the computing infrastructure essential to our analyses. Finally, we acknowledge the enduring support for the construction and operation of the LHC and the CMS detector provided by the following funding agencies: BMBWF and FWF (Austria); FNRS and FWO (Belgium); CNPq, CAPES, FAPERJ, FAPERGS, and FAPESP (Brazil); MES (Bulgaria); CERN; CAS, MoST, and NSFC (China); COLCIENCIAS (Colombia); MSES and CSF (Croatia); RPF (Cyprus); SENESCYT (Ecuador); MoER, ERC IUT, PUT and ERDF (Estonia); Academy of Finland, MEC, and HIP (Finland); CEA and CNRS/IN2P3 (France); BMBF, DFG, and HGF (Germany); GSRT (Greece); NKFIA (Hungary); DAE and DST (India); IPM (Iran); SFI (Ireland); INFN (Italy); MSIP and NRF (Republic of Korea); MES (Latvia); LAS (Lithuania); MOE and UM (Malaysia); BUAP, CINVESTAV, CONACYT, LNS, SEP, and UASLP-FAI (Mexico); MOS (Montenegro); MBIE (New Zealand); PAEC (Pakistan); MSHE and NSC (Poland); FCT (Portugal); JINR (Dubna); MON, RosAtom, RAS, RFBR, and NRC KI (Russia); MESTD (Serbia); SEIDI, CPAN, PCTI, and FEDER (Spain); MOSTR (Sri Lanka); Swiss Funding Agencies (Switzerland); MST (Taipei); ThEPCenter, IPST, STAR, and NSTDA (Thailand); TUBITAK and TAEK (Turkey); NASU (Ukraine); STFC (United Kingdom); DOE and NSF (USA).

\hyphenation{Rachada-pisek} Individuals have received support from the Marie-Curie program and the European Research Council and Horizon 2020 Grant, contract Nos.\ 675440, 752730, and 765710 (European Union); the Leventis Foundation; the A.P.\ Sloan Foundation; the Alexander von Humboldt Foundation; the Belgian Federal Science Policy Office; the Fonds pour la Formation \`a la Recherche dans l'Industrie et dans l'Agriculture (FRIA-Belgium); the Agentschap voor Innovatie door Wetenschap en Technologie (IWT-Belgium); the F.R.S.-FNRS and FWO (Belgium) under the ``Excellence of Science -- EOS" -- be.h project n.\ 30820817; the Beijing Municipal Science \& Technology Commission, No. Z181100004218003; the Ministry of Education, Youth and Sports (MEYS) of the Czech Republic; the Lend\"ulet (``Momentum") Program and the J\'anos Bolyai Research Scholarship of the Hungarian Academy of Sciences, the New National Excellence Program \'UNKP, the NKFIA research grants 123842, 123959, 124845, 124850, 125105, 128713, 128786, and 129058 (Hungary); the Council of Science and Industrial Research, India; the HOMING PLUS program of the Foundation for Polish Science, cofinanced from European Union, Regional Development Fund, the Mobility Plus program of the Ministry of Science and Higher Education, the National Science Center (Poland), contracts Harmonia 2014/14/M/ST2/00428, Opus 2014/13/B/ST2/02543, 2014/15/B/ST2/03998, and 2015/19/B/ST2/02861, Sonata-bis 2012/07/E/ST2/01406; the National Priorities Research Program by Qatar National Research Fund; the Ministry of Science and Education, grant no. 3.2989.2017 (Russia); the Programa Estatal de Fomento de la Investigaci{\'o}n Cient{\'i}fica y T{\'e}cnica de Excelencia Mar\'{\i}a de Maeztu, grant MDM-2015-0509 and the Programa Severo Ochoa del Principado de Asturias; the Thalis and Aristeia programs cofinanced by EU-ESF and the Greek NSRF; the Rachadapisek Sompot Fund for Postdoctoral Fellowship, Chulalongkorn University and the Chulalongkorn Academic into Its 2nd Century Project Advancement Project (Thailand); the Nvidia Corporation; the Welch Foundation, contract C-1845; and the Weston Havens Foundation (USA).
\end{acknowledgments}

\bibliography{auto_generated}
\cleardoublepage \appendix\section{The CMS Collaboration \label{app:collab}}\begin{sloppypar}\hyphenpenalty=5000\widowpenalty=500\clubpenalty=5000\vskip\cmsinstskip
\textbf{Yerevan Physics Institute, Yerevan, Armenia}\\*[0pt]
A.M.~Sirunyan$^{\textrm{\dag}}$, A.~Tumasyan
\vskip\cmsinstskip
\textbf{Institut f\"{u}r Hochenergiephysik, Wien, Austria}\\*[0pt]
W.~Adam, F.~Ambrogi, T.~Bergauer, M.~Dragicevic, J.~Er\"{o}, A.~Escalante~Del~Valle, M.~Flechl, R.~Fr\"{u}hwirth\cmsAuthorMark{1}, M.~Jeitler\cmsAuthorMark{1}, N.~Krammer, I.~Kr\"{a}tschmer, D.~Liko, T.~Madlener, I.~Mikulec, N.~Rad, J.~Schieck\cmsAuthorMark{1}, R.~Sch\"{o}fbeck, M.~Spanring, D.~Spitzbart, W.~Waltenberger, C.-E.~Wulz\cmsAuthorMark{1}, M.~Zarucki
% ...
\vskip\cmsinstskip
\dag: Deceased\\
1:  Also at Vienna University of Technology, Vienna, Austria\\
% ...
\end{sloppypar}
%%% END EDITABLE REGION %%%
\end{document}